%% file: cam-iii.tex
\documentclass[10pt,a4paper,twoside,dvipdfm]{article}
\usepackage{epsfig}
\usepackage{baltlat6}
\usepackage{array}
\usepackage{wrapfig}
\usepackage{longtable}
\usepackage{lscape}
\begin{document}
\ \
\vspace{0.5mm}
\setcounter{page}{1}
\vspace{8mm}

\titlehead{Baltic Astronomy, vol.\,17, 1--19, 2008}

\titleb{YOUNG STARS IN THE CAMELOPARDALIS DUST AND\\ MOLECULAR CLOUDS.
III. THE GL\,490 REGION}

\begin{authorl}
\authorb{V. Strai\v zys}{} and
\authorb{V. Laugalys}{}
\end{authorl}

\moveright-3.2mm
\vbox{
\begin{addressl}
\addressb{}{Institute of Theoretical Physics and Astronomy, Vilnius
University,\\  Go\v stauto 12, Vilnius LT-01108, Lithuania}
\end{addressl}
}

\submitb{Received 2007 November 2; accepted 2007 November 30}

\begin{summary} Using the infrared photometry data extracted from the
2MASS, IRAS and MSX databases, 50 suspected young stellar objects (YSOs)
are selected from about 37\,500 infrared objects in the
3\degr\,$\times$\,3\degr\ area with the center at $\ell$, $b$ =
142.5\degr, +1.0\degr, in the vicinity of the young stellar object
GL\,490 in the dark cloud DoH 942 (Dobashi et al. 2005).  The spectral
energy distributions between 700 nm and 100 $\mu$m suggest that most of
the selected objects may be YSOs of classes I and II.  In the
color-magnitude diagram $K_s$ vs.  $H$--$K_s$ the suspected YSOs occupy
an area right of the main sequence what can be interpreted as being
caused by the effects of luminosity, interstellar and circumstellar
reddening and infrared thermal emission in circumstellar envelopes and
disks.  \end{summary}


\resthead{Young stars in the Camelopardalis dust and molecular clouds.
III.}{V. Strai\v zys, V. Laugalys}

\sectionb{1}{INTRODUCTION}

In the previous papers (Strai\v zys \& Laugalys 2007a,b, Papers I and
II) we made census of young stellar objects in the Camelopardalis
segment of the Local spiral arm ($\ell$, $b$\,= 132--158\degr,
$\pm$\,12\degr).  More than 40 stars of the Cam OB1 association, about
20 young stars of lower masses exhibiting emission in H$\alpha$ or
belonging to irregular variable stars of types IN and IS, as well as 42
infrared young stellar objects (YSOs) in the Local arm were identified.
Among the latter objects, the most prominent is a high-mass young object
GL\,490, embedded in the densest part of the dust cloud DoH\,942
(Dobashi et al. 2005).  \footnote{~The clouds in the Dobashi et al.
(2005) high-resolution atlas in Paper I  were named Tokyo
clouds.} All of the identified YSOs have $H$--$K_s$\,$\geq$\,1.0, since
bluer objects were difficult to identify among thousands of objects
having no relation to star forming.

Trying to find more YSOs in the area, we reduced the limiting $H$--$K_s$
from 1.0 to 0.75, decreasing at the same time the size of the
investigated area down to 3\degr\,$\times$\,3\degr.  In the area
centered at $\ell$, $b$ = 142.5\degr, +1.0\degr\ we have analyzed the
infrared objects measured in the 2MASS, IRAS and MSX surveys.  For IRAS
objects the selection limit of $H$--$K_s$ was decreased from 0.75 to
0.5.

For a comparison, the infrared objects were also considered in the
standard area of the same size centered at $\ell$, $b$ = 144.0\degr,
+3.5\degr, in a relatively transparent direction.  Both areas are shown
in Figure 1, with dust clouds from the Dobashi et al.  (2005) atlas
shown in the background.


\begin{figure}[!t]
\centerline{\psfig{figure=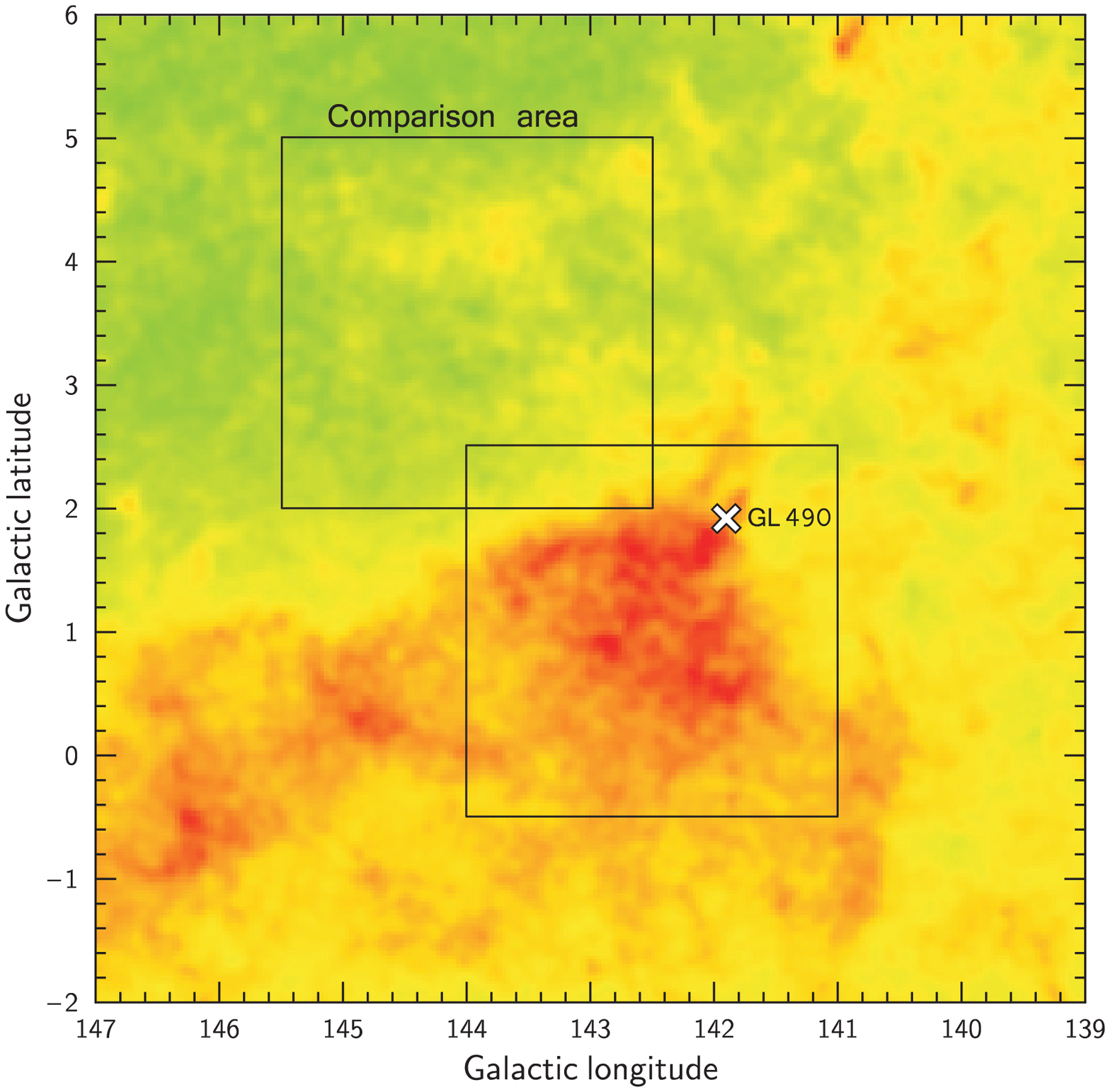,width=80mm,angle=0,clip=true}}
\vspace{0mm}
\captionb{1}{Positions of the GL\,490 area and the comparison area in
Galactic coordinates. In the background dust clouds from the Dobashi et
al. (2005) atlas are shown. The position of YSO GL\,490 in the
corner of the DoH\,942 cloud is shown as the white cross.}
\end{figure}

\sectionb{2}{IDENTIFICATION OF THE PRE-MAIN-SEQUENCE OBJECTS}

Figure 2 shows the $J$--$H$ vs.  $H$--$K_s$ diagram for about 37\,500
stars measured in the 2MASS survey with the errors $\leq$\,0.05 mag
(Cutri et al. 2003; Skrutskie et al. 2006).  The picture is quite
similar to that shown in Fig.\,1 of Paper II.  In the comet-like
crowding of dots the orange line designates the intrinsic main sequence,
the yellow line K--M giants and the blue line the intrinsic locus of T
Tauri-type stars from Meyer et al.  (1997).  In both figures the `comet
head' is composed mostly of normal stars of different spectral classes
with small interstellar extinction.  The upper `tail' is composed of
normal heavily reddened background stars, mostly of K and M giants.


\begin{figure}[!t]
\centerline{\psfig{figure=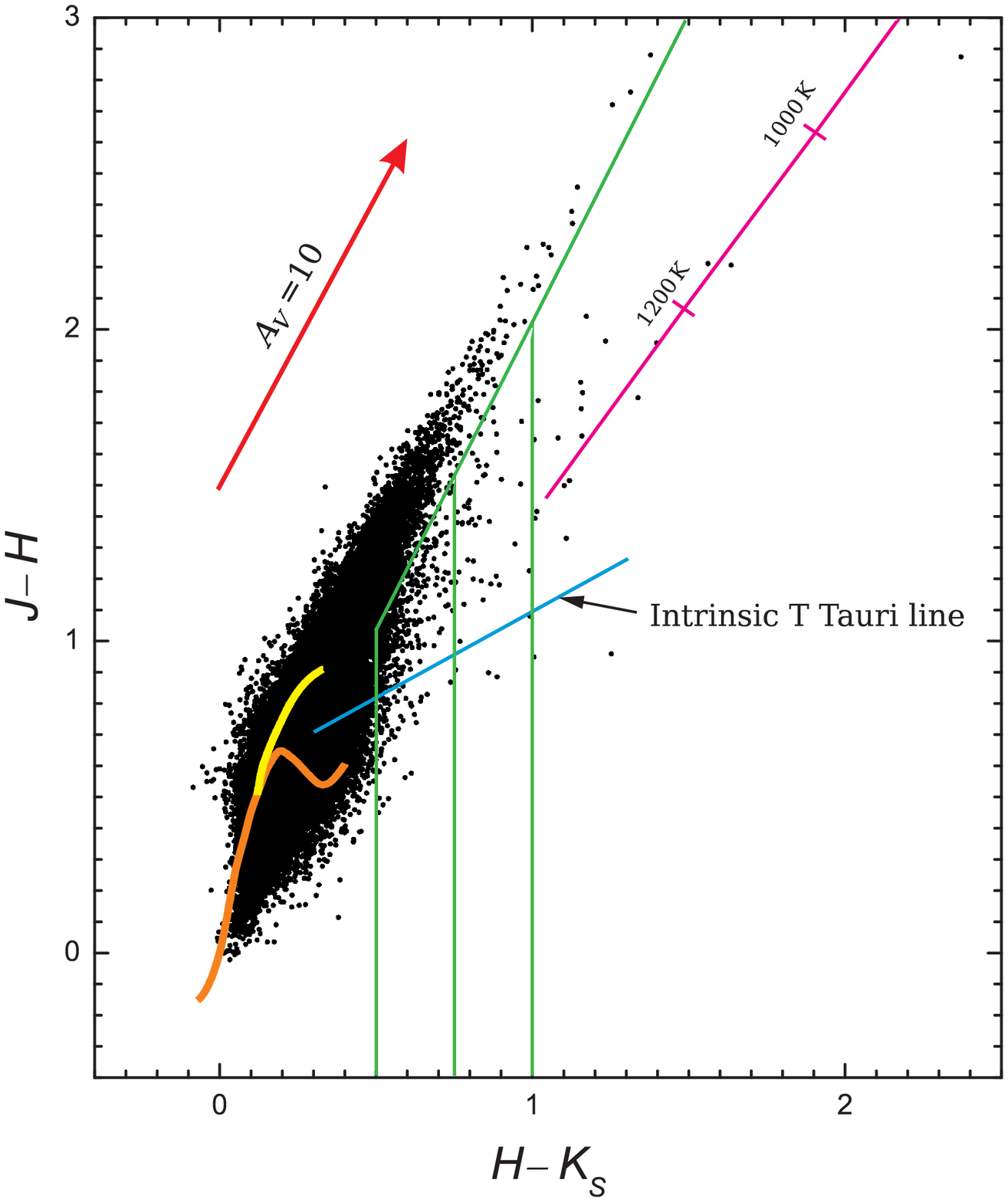,width=80mm,angle=0,clip=true}}
\vspace{0mm}
\captionb{2}{ The $J$--$H$ vs.  $H$--$K_s$ diagram for
37\,500 objects in the GL\,490 area.  The intrinsic main-sequence
and K--M giant lines are shown in orange and yellow, respectively.  The
blue line designates the intrinsic locus of T Tauri stars, the violet
line is the locus of black bodies.  The length of the reddening vector
(shown in red) corresponds to the extinction in the $V$ passband of 10
mag.  The three green vertical lines and the reddening line running
through the point $J$--$H$ = $H$--$K_s$ = 0 isolate the regions where
the presence of young stellar objects was investigated (see the text). }
\end{figure}

As it was shown in Paper II, the lower `tail', running more or less
along the black-body line, contains YSOs of different masses and
evolutionary stages.  In this part of the diagram, along with the young
objects, we expect to find also M-type giants of the latest subclasses,
including oxygen-rich and carbon-rich long-period variables, OH/IR
stars, carbon-rich stars of spectral type N, Be stars, infrared dusty
galaxies and quasars.  Since {\em JHK} photometry is not sufficient for
the identification of young stars, either spectroscopic or infrared
photometric observations at longer wavelengths are essential.

Suspected YSOs in Figure 2 were isolated in the box limited by the
three lines: from the left and right sides by two vertical lines at
$H$--$K_s$\,=\,0.75 and 1.0, and from the top by the interstellar
reddening line corresponding to
$$
Q_{JHK} = (J - H) - 1.85 (H - K_s) = 0.0.
$$
As was shown in Paper II this condition excludes the majority of normal
stars of various temperatures, luminosities and reddenings (except
heavily reddened O--B stars and the coolest M giants and dwarfs). The
objects with $H$--$K_s$\,$\geq$\,1.0 were considered in Paper II.  In
the box between $H$--$K_s$ = 0.75 and 1.0 we found 37 objects.

Apart from YSOs, in this part of the diagram we expect to find reddened
M-type giants of the latest subclasses, AGB OH/IR stars, carbon-rich
stars of spectral type N, Be stars, dusty spiral galaxies and quasars.
In our sample of 37 objects we identified, using the Simbad database,
two carbon stars, one galaxy and two radio sources.  The remaining 32
objects are listed in Table 1, continuing the same sequential numeration
as in Table 1 of Paper II.  As earlier, magnitudes, color indices and
$Q$-parameters are rounded to two decimal places.  Seven objects


\input table1.tex


\input table2.tex

\noindent of Table 1 have been measured by IRAS and three by MSX. These
objects are listed in Table 2, together with five objects in the same
area (SL\,89, SL\,93, SL\,95, SL\,102 and SL\,107) identified in Paper
II as YSOs with $H$--$K_s$\,$\geq$\,1.0.

Additionally, in Tables 1 and 2 we list 13 objects with $H$--$K_s$
between 0.5 and 0.75 measured by IRAS (numbers SL\,175--187). These
objects are also YSO candidates (see Section 6). However, the
identification of IRAS and 2MASS sources in some cases is problematic
due to low accuracy of the IRAS coordinates.


\begin{figure}[!t]
\centerline{\psfig{figure=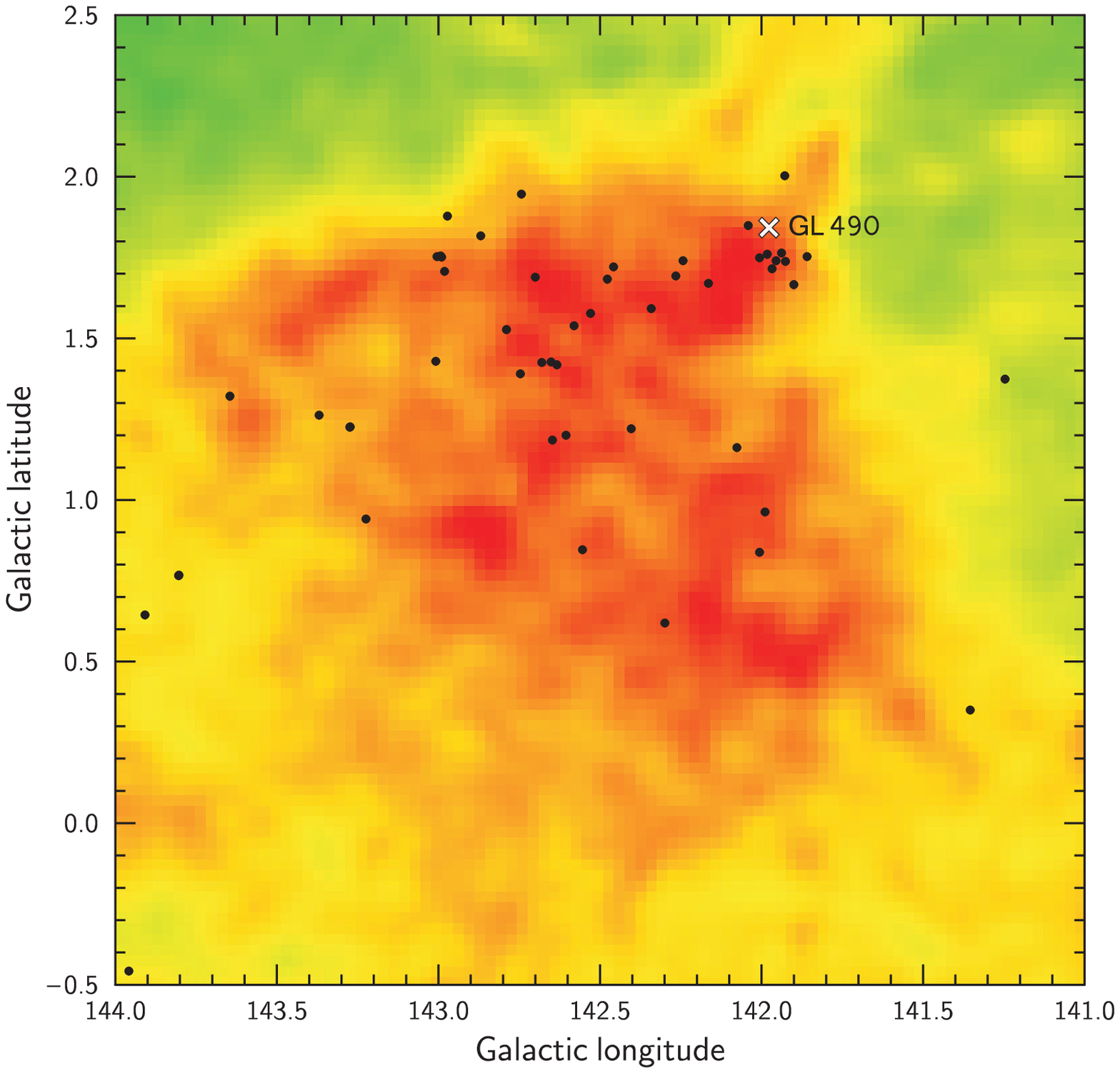,width=86mm,angle=0,clip=true}}
\vspace{0mm}
\captionb{3}{Positions of the suspected YSOs in the GL\,490 area
in the Galactic coordinates. Dust clouds from the Dobashi et al. (2005)
atlas are shown in the background. The position of the YSO GL\,490
 is shown by the white cross.}
\vskip4mm
\centerline{\psfig{figure=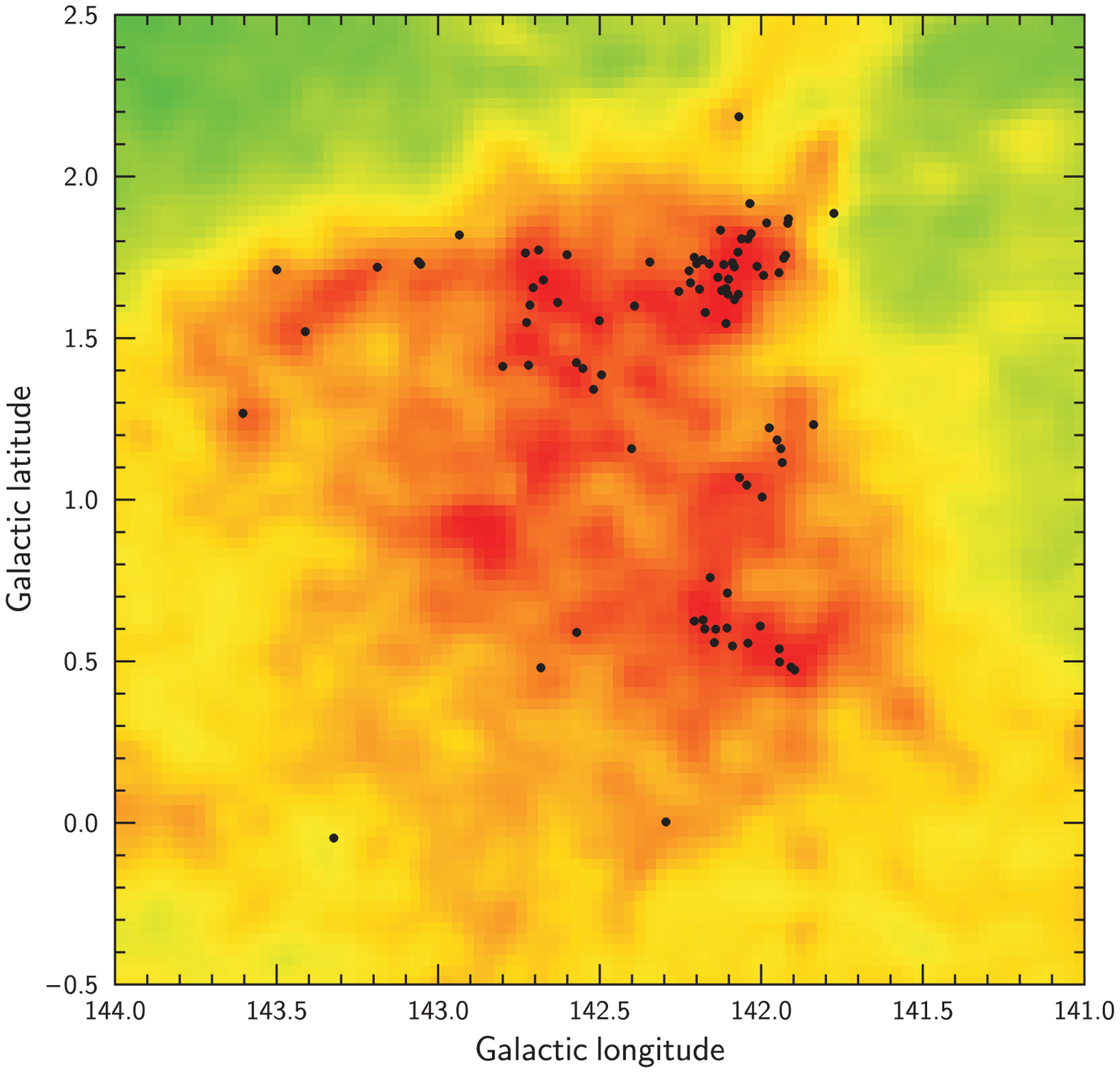,width=86mm,angle=0,clip=true}}
\vspace{0mm}
\captionb{4}{Positions of the suspected K--M giants in the GL\,490 area.
Dust clouds are the same as in Figure 3.}
\end{figure}

In the Galactic coordinates (Figure 3) we plot 45 objects from Table 1
plus five objects from Paper II listed above.  In the background, dust
clouds from the Dobashi et al.  (2005) atlas are shown.  It is evident
that the suspected YSOs concentrate in the densest clump of the DoH\,942
cloud:  11 objects are crowded at $\ell$, $b$\,=\,141.9--142.1\degr,
+1.7 -- +2.0\degr, around the position of the high-mass YSO, GL\,490.
Hodapp (1990, 1994) described a smaller cluster at GL\,490 seen on his
$K$-band images.  However (see the next section), clustering of infrared
objects in the direction of a dense dust cloud is not necessary related
to young objects.

\sectionb{3}{DISTRIBUTION OF REDDENED K--M GIANTS IN THE AREA}

With the aim to compare the surface distributions of the suspected YSOs
and the reddened stars of normal spectral classes, we have separated 88
stars of the upper `tail' in Figure 2 satisfying the following
conditions:  (1) $H$--$K_s$\,$\geq$\,0.75 and (2) $Q_{JHK}$\,$>$\,0.2.
These stars are listed in Table 3. In two-color diagram these stars are
dispersed along the reddening line of K and M giants, therefore we use
the letters K-M as the prefix to their numbers.  Most probably they are
field stars located behind the DoH\,942 cloud, with the extinction $A_V$
between 10 and 20 mag.  K and M giants in the $K$ passband are quite
bright:  absolute magnitudes $M_K$ of K0\,III and M5\,III stars are
--1.6 mag and --6.3 mag, respectively.  Such stars appear in our sample
even with heavy extinction and being located at distances of a few kpc.
A simple calculation shows that at a distance of 1 kpc, where the
DoH\,942 cloud is supposed to lie, all main-sequence stars cooler than
F5\,V should be fainter than our limiting magnitude $K$ = 12.5.  Stars
of spectral classes B and A should be also absent in the sample:  they
are excluded by condition (2) listed above, since their $Q_{JHK}$ is
close to zero.

We could not find any IRAS source among the selected 88 objects of the
upper `tail'.  Only six sources were identified with MSX. This is an
additional argument that these objects are normal K--M stars without
dust envelopes.  In the IRAS passbands they were too faint to be
measured.

\vspace{-4mm}


\input table3

Figure 4 exhibits the surface distribution of the supposed heavily
reddened K--M giants in the Galactic coordinates.  One can find strong
evidence that these stars have a tendency to concentrate in the
direction of the densest dust clouds, showing similarity to the
distribution of possible YSOs (Figure 3).  However, K--M giants are
distributed broader, they are not so concentrated to the GL 490 cloud as
YSOs.  We should expect that K--M giants in the background populate all
the area with more or less uniform surface density, and many of them are
present in the $J$--$H$ vs.  $H$-$K_s$ diagram of Figure 2. However,
their color indices $H$--$K_s$ are $<$\,0.75, consequently, they do not
appear in Figure 4.

The tendency of both the YSOs and the heavily reddened K--M giants to
concentrate apparently in the direction of dust clouds prevents using
the clustering factor alone as approval of physical relation between the
stars and the cloud.  This can lead to misinterpretation of distant red
giants as a cluster of infrared YSOs.  However, this ambiguity can be
avoided with {\it JHK} photometry at hand since K--M giants and YSOs are
located in different `tails' in the $J$--$H$ vs.  $H$--$K_s$ diagram.

\sectionb{5}{THE COMPARISON AREA}

For a better understanding of the reliability of identification of YSOs
and K--M giants we decided to apply the same procedure for an area
located outside the dust cloud but close to the GL 490 area.  We expect
that such an area should contain almost the same amount of background
K--M giants, AGB stars, galaxies and quasars which are the main YSO
simulators.  For this aim we selected the 3\,$\times$\,3\degr\ area
centered at $\ell$, $b$ = 144.0\degr, +3.5\degr\ which is shown in
Figure 1.


\begin{figure}[!t]
\centerline{\psfig{figure=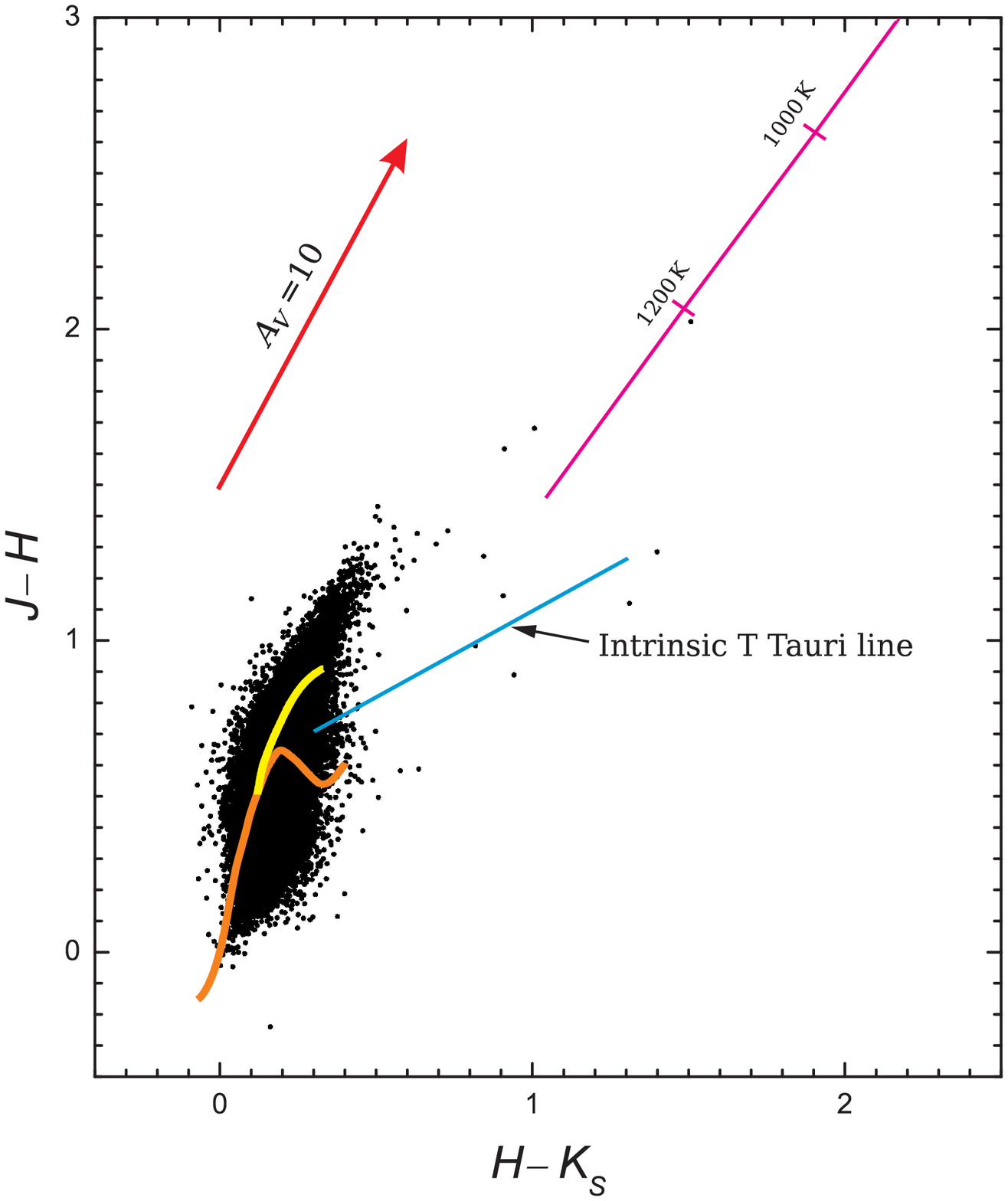,width=90mm,angle=0,clip=true}}
\vspace{0mm}
\captionb{5}{ The $J$--$H$ vs.  $H$--$K_s$ diagram for
35\,200 objects in the comparison area shown in Figure 1.  The intrinsic
main-sequence and K--M giant lines, the intrinsic locus of T Tauri stars,
the locus of black bodies and the reddening vector are the same as in
Figure~2.}
\end{figure}

The $J$--$H$ vs.  $H$--$K_s$ diagram for 35\,200 infrared objects,
selected in the comparison area, is shown in Figure 5. Notice that the
numbers of objects in both the GL\,490 area and the comparison area are
similar.  However, their distribution in the two-color diagram is quite
different.  The upper `tail' of K--M giants for the comparison area is
quite short -- the interstellar extinction $A_V$ of the most reddened
background stars does not exceed 4--5 mag.  The lower `tail', observed
in the GL\,490 area, is amost absent:  we find only 10 stars near the
intrinsic line of T Tauri stars and near the black-body line with
$H$--$K_s$\,$>$\,0.75.  Six of them are IRAS and MSX objects, among them
four objects are carbon stars and one object is a Mira variable (LL
Cam).

\sectionb{6}{SPECTRAL ENERGY DISTRIBUTIONS}

The best way to discriminate between YSOs and old AGB objects (Mira
variables, N-type carbon stars and OH/IR objects) is to construct
infrared spectral energy distribution (SED) curves using the 2MASS, IRAS
and MSX data.  The 2MASS data alone are not sufficient since all these
types of stars occupy the same area in the $J$--$H$ vs.  $H$--$K_s$
diagram.

Table 2 of the present paper lists 25 objects in the vicinity of GL\,490
which have reliable fluxes at least in one of the four IRAS passbands
and/or have reliable MSX fluxes in the 8.3 $\mu$m passband.  Their
red $F$ magnitudes at 710 nm were selected from the GSC 2.2 catalog.
For all these objects SEDs were calculated as described in Paper II and
for 12 of them are plotted in Figures 6 and 7. Almost all these objects
exhibit strong infrared excesses at $\lambda$\,$>$\,2.2 nm, i.e.,
probably they are pre-main-sequence objects in different evolutionary
stages (Lada 1987; Robitaille


\vbox{
\hbox{\parbox[t]{85mm}{\psfig{figure=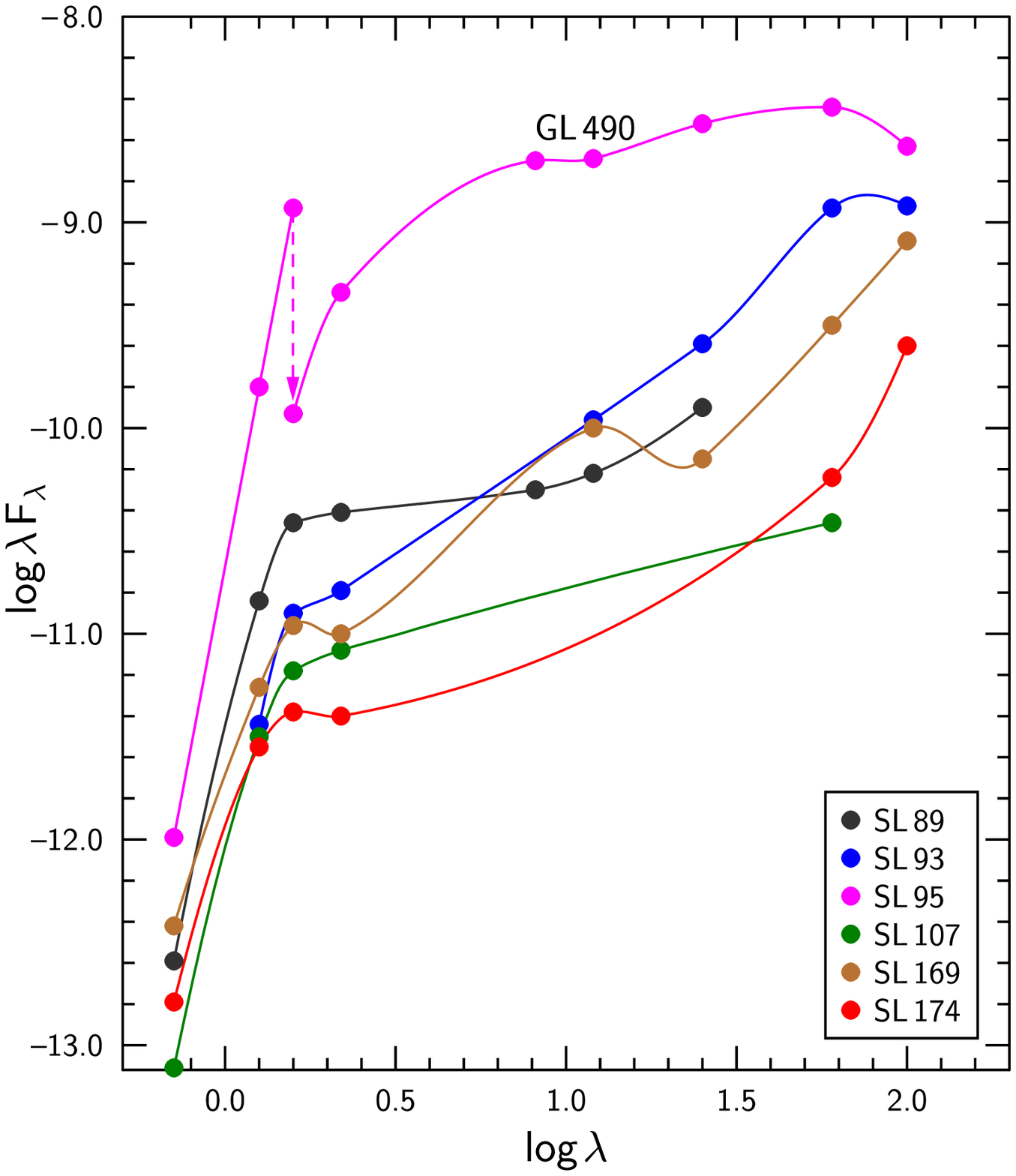,width=80mm,angle=0,clip=true}}
\parbox[t]{37mm}{\vskip-2.5cm
\captionr{6}{Spectral energy distributions for six objects of Table 2
which are most similar to the Class I YSOs.}}}
\vskip5mm
\hbox{\parbox[t]{85mm}{\psfig{figure=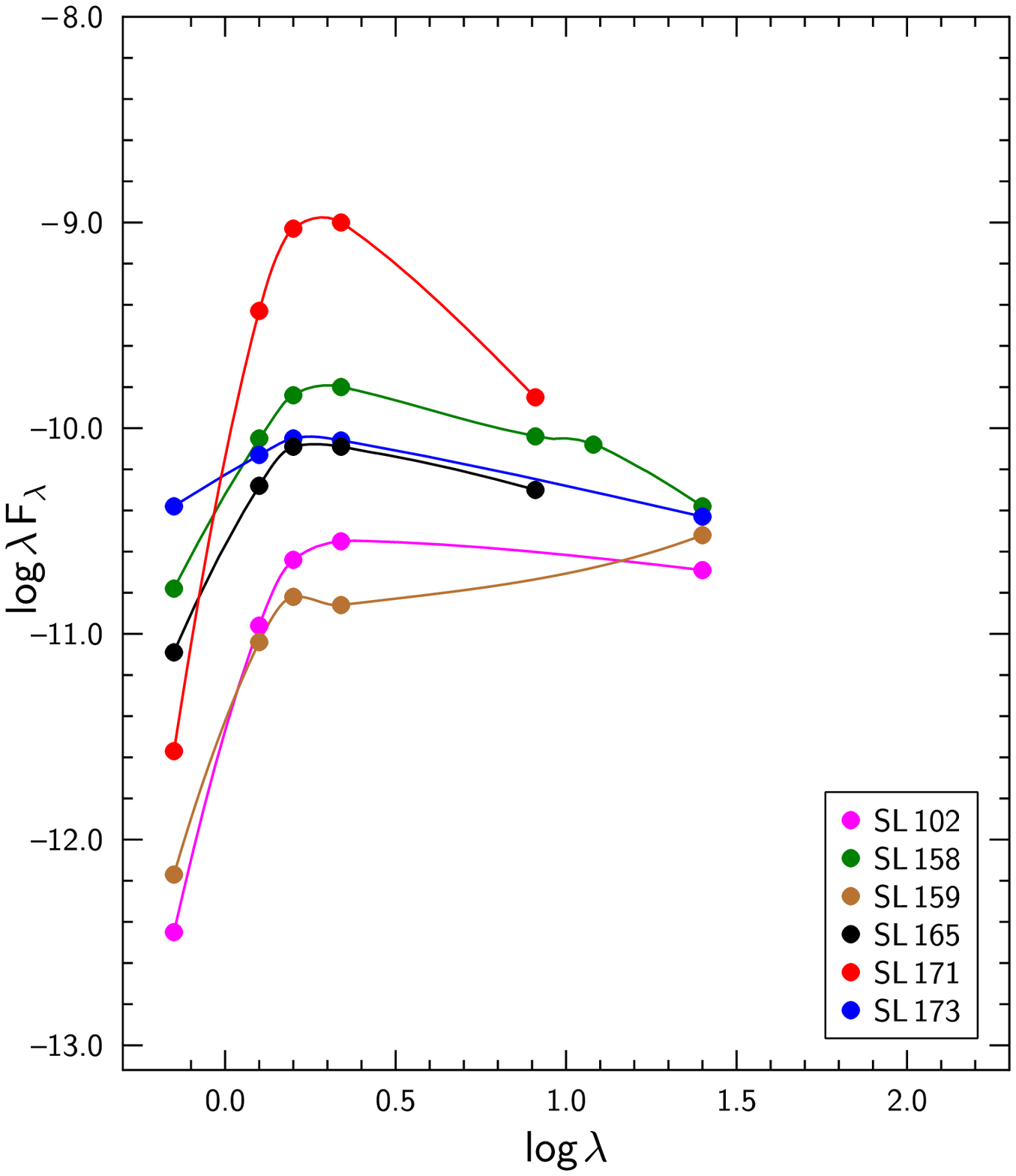,width=80mm,angle=0,clip=true}}
\parbox[t]{37mm}{\vskip-2.5cm
\captionr{7}{Spectral energy distributions for six objects of Table 2
which are most similar to the Class II YSOs.}}}
}

\noindent  et al. 2006).  Sixteen objects exhibit SEDs which are similar
to Class I with a steep rise of the flux in the far infrared, eight are
Class II objects with the flat flux, and one object (SL\,171) may be a
heavily reddened Herbig Ae/Be star or a distant AGB object.

Among the suspected K--M giants (upper `tail'), only five objects were
measured by MSX and no objects by IRAS (Table 4).  SEDs of these stars
are plotted in Figure 8. There is no doubt that all of them are similar
to heavily reddened K--M stars without infrared excesses (compare this
figure with Fig.~5 in Paper II).  This confirms our claim that the upper
`tail' in the $J$--$H$ vs.  $H$--$K_s$ diagram (Figure 1) is formed
mainly by normal late-type giants.


\begin{figure}[!th]
\hbox{\parbox[t]{85mm}{\psfig{figure=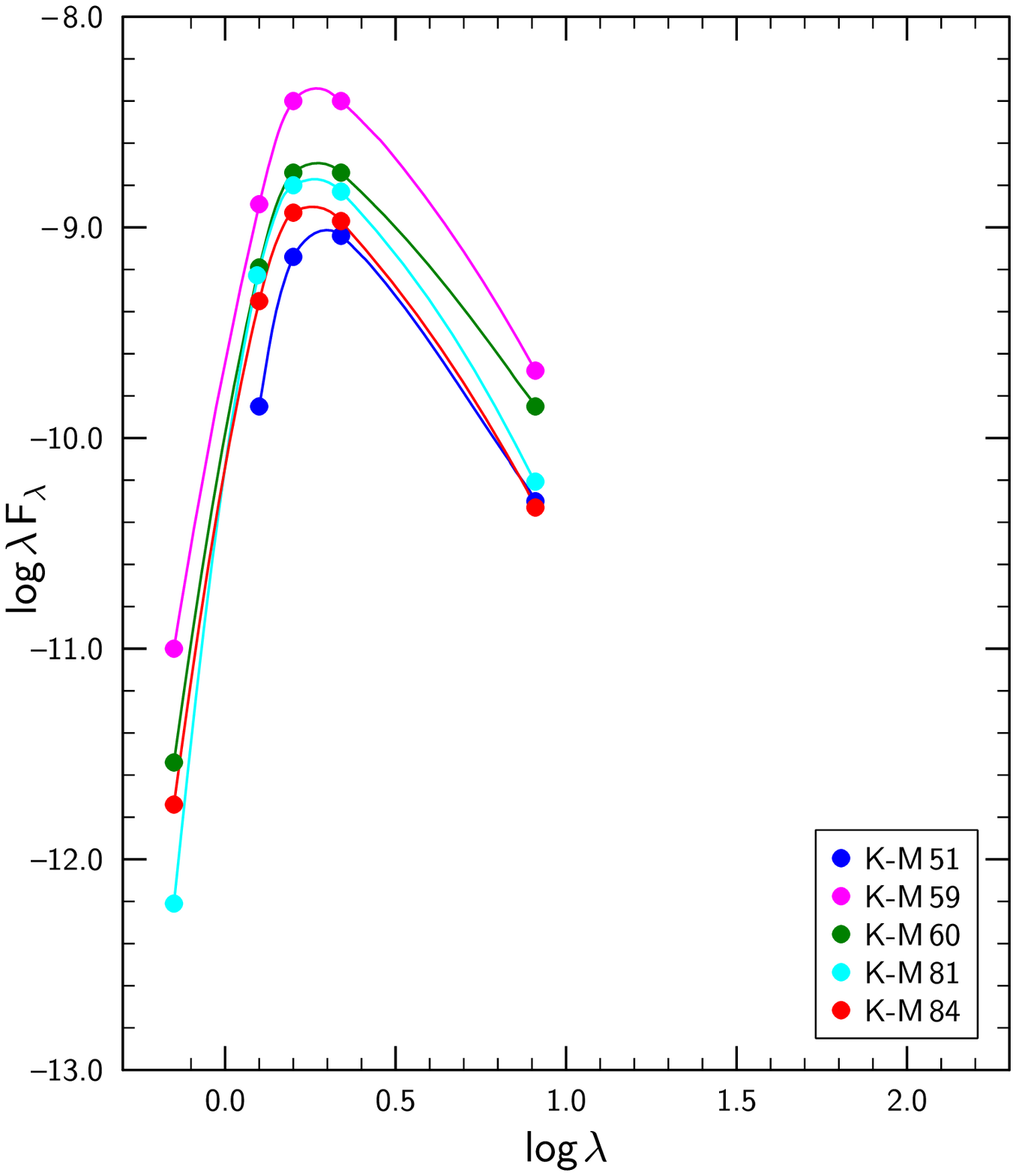,width=80mm,angle=0,clip=true}}
\parbox[t]{37mm}{\vskip-2.5cm\captionr{8}{\hbox{Spectral energy}
distributions for five stars, suspected heavily reddened K--M giants.}}}
\end{figure}


\begin{figure}[!th]
\centerline{\psfig{figure=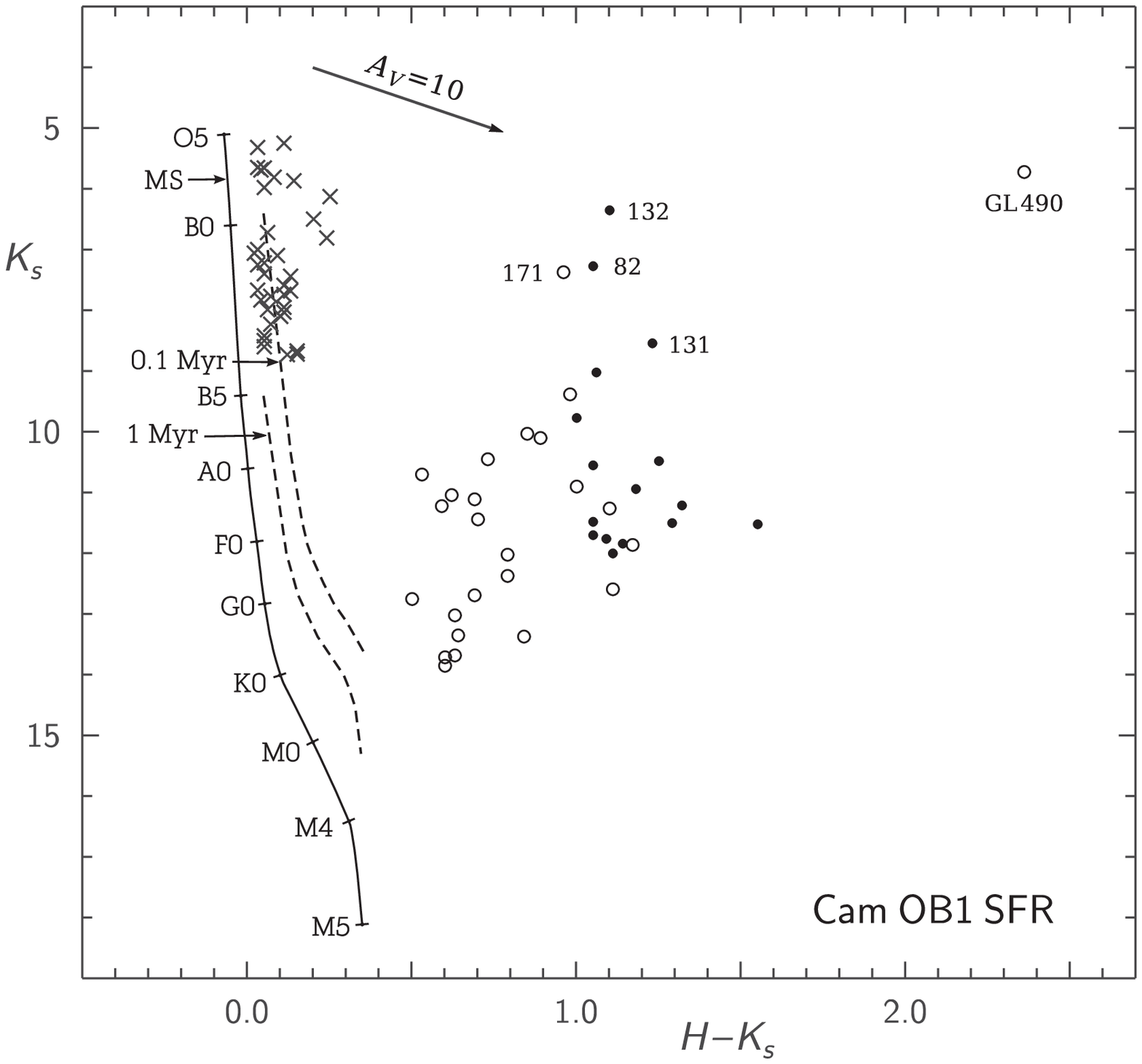,width=100mm,angle=0,clip=true}}
\vspace{.8mm}
\captionb{9}{The $K_s$ vs.  $H$--$K_s$ diagram for the Cam OB1
star-forming region located at a distance of 900 pc.  Open circles are
YSOs from the GL\,490 area of 3\degr\,$\times$\,3\degr\ size, dots are
YSOs from the Cam OB1 association area of 26\degr\,$\times$\,24\degr\
size. Crosses are the Cam OB1 association stars of spectral classes O--B.  The
main-sequence and isochrones for 0.1 Myr and 1.0 Myr are also plotted.}
\end{figure}


\begin{figure}[!th]
\vspace{3mm}
\centerline{\psfig{figure=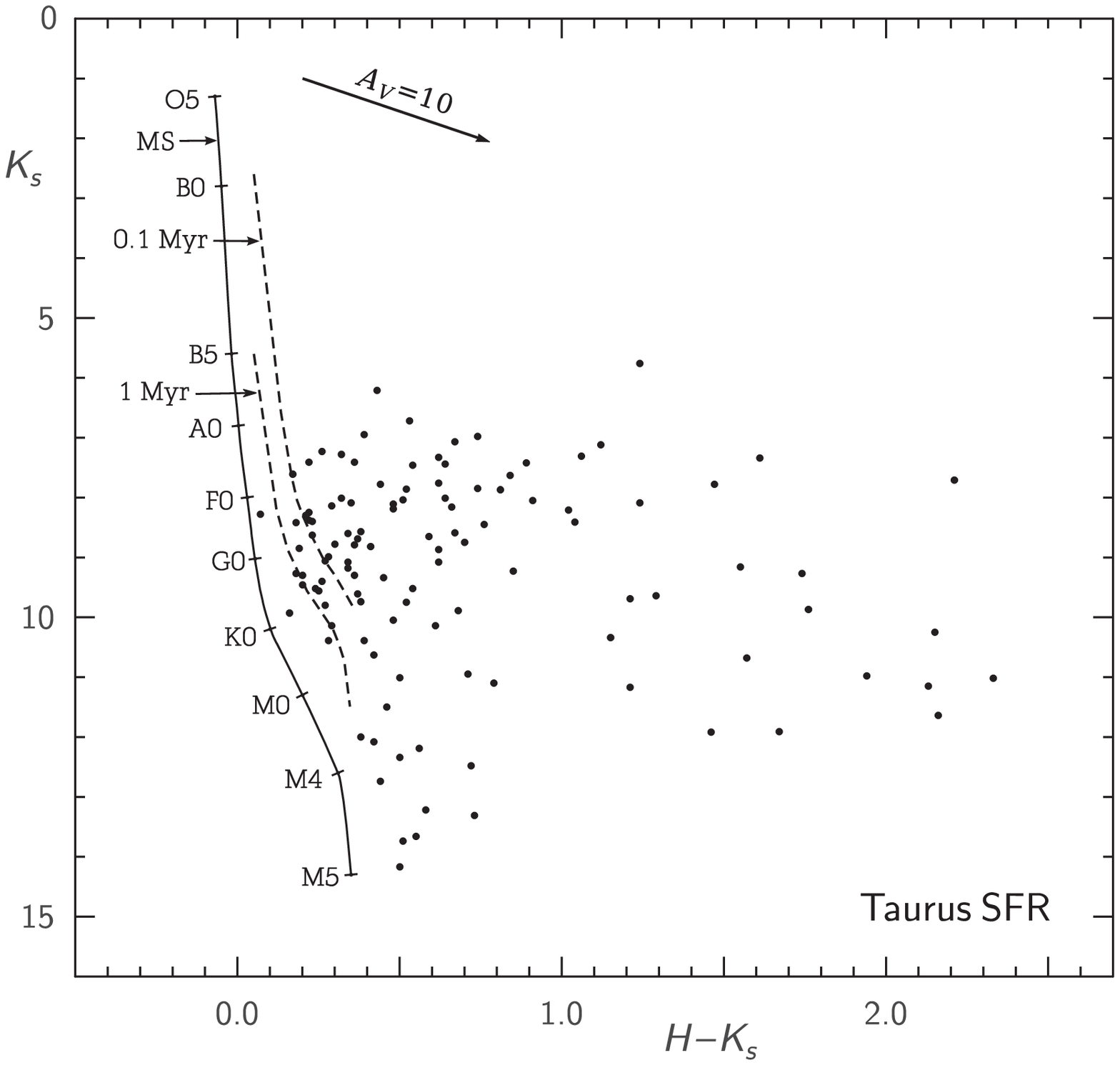,width=100mm,angle=0,clip=true}}
\vspace{.8mm}
\captionb{10}{The $K_s$ vs. $H$--$K_s$ diagram for YSOs of the Taurus
star-forming region located at a distance of 150 pc. The data are from
Brice\~{n}o et al. (2002).}
\end{figure}

\vskip1mm


\input table4.tex

\newpage

\sectionb{7}{NEAR INFRARED COLOR-MAGNITUDE DIAGRAM FOR THE\\ CAM~OB1
STAR-FORMING REGION}
\vskip1mm

More information about YSOs can be obtained from the infrared color vs.
magnitude diagrams.  One of these, relating the $K_s$ magnitude with the
$H$--$K_s$ color index, is presented in Figure 9 for YSOs belonging to
the Cam OB1 SFR.  Only the objects, whose dependence to YSOs has been
confirmed by IRAS and/or MSX photometry, are plotted.  Open circles
designate 25 YSOs from the GL\,490 3\,$\times$\,3\degr\ area (Table 2)
and dots designate 16 YSOs belonging to the Cam OB1 SFR in the
26\degr\,$\times$\,24\degr\ area from Paper II.  The unreddened main
sequence corresponds to a distance of 900 pc.  The isochrones for the
ages 0.1 and 1.0 Myr are from Hillenbrand \& Carpenter (2000).  In the
color-magnitude diagram, along with the YSOs, we also plot O--B3 stars
of the association (crosses).  All of them are concentrated near the
zero value of $H$--$K_s$.  Some of B-stars are of luminosity classes IV
and III, and they overlap the reddened O-type stars.

The distribution of YSOs in Figure 9 is the result of combined effects
of variable masses, temperatures, ages, near infrared excesses (due to
envelopes and disks) and extinctions (both interstellar and
circumstellar).  The additional scatter of objects is introduced by
various orientations of envelope cavities for Stage I objects and rings
for Stage II objects (see Robitaille et al. 2006, 2007).  Sometimes, the
magnitude of YSO is influenced by additional radiation of the
surrounding reflection or emission nebula, consequently, the result
depends on the aperture used.  At the present stage of investigation we
have no possibility to disentangle all these effects for individual
objects.

The positions of YSOs in the $K_s$ vs. $H$--$K_s$ diagram above the main
sequence can be explained by the following effects.

$\bullet$ The shift upward due to a larger diameter and luminosity:  for
a typical YSO age of 0.1 Myr the isochrone is above the main sequence
by $\sim$\,3 mag.

$\bullet$ The shift upward and to the right due to near-infrared thermal
emission in the circumstellar envelope or disk.  According to
Hillenbrand \& Carpenter (2000) for the Taurus YSOs the $K$ excess is
from zero to $\sim$\,1.5 mag and the $H$--$K$ excess is from zero to
$\sim$\,0.8 mag, both excesses are linearly correlated.

$\bullet$ The shift down and to the right along the reddening lines due
to extinction and reddening by the interstellar and circumstellar dust.
The slope of the reddening line is $A_K$/$E_{H-K}$ = 1.79 (Bessell \&
Brett 1988), and the ratio of extinctions is $A_K$/$A_V$ = 1.14
(Cardelli et al. 1989). These ratios for the dust in the circumstellar
envelope and  disk can be somewhat different.

Probably, the best criterion for identifying YSOs of Stage I is their
faintness in the visual wavelengths.  In Figure 9 we find objects which
are sufficiently bright in the $J$, $H$ and $K_s$ passbands but are
invisible in the Palomar DSS2 red and blue plates.  These stars can be
identified in Table 1 by the absence of magnitude in the $F$ column.
Such stars are also frequent in the Taurus clouds, despite their
proximity to the Sun.  Probably, this is caused by circumstellar dust
shells and disks giving the extinctions $A_V$ of the order of 30 mag and
more (Myers et al. 1987; Whitney et al. 2003).  However, if the object
is directed to us by its envelope cavity or we are viewing along the
disk axis, the circumstellar extinction is much smaller.

For comparison, the $K$ vs.  $H$--$K_s$ diagram for YSOs in the Taurus
SFR is presented in Figure 10, taking the $K_s$ and $H$--$K_s$ data from
Brice\~{n}o et al.  (2002) and a distance of 150 pc.  The
3\,$\times$\,3\degr\ area near GL\,490 at a distance of 900 pc
corresponds to 17\,$\times$\,17\degr\ at a distance of the Taurus SFR.
This means that the real sizes of the GL\,490 area and the Taurus SFR
are more or less equal.  Although the general appearance of the
color-magnitude diagrams of the Cam OB1 and Taurus star-forming regions
is similar, there are appreciable differences between them.  We should
not pay attention to the number differences of YSOs in both regions
since the identification of YSOs in the Cam OB1 area (open circles in
Figure 9) is affected by selection effects, such as the accuracy limit
of 2MASS photometry, the presence or absence of the IRAS and MSX data,
the limiting magnitude due to large distance, etc.

However, some differences are obvious.  For example, the Cam OB1 SFR
contains massive YSOs which are absent in the Taurus SFR.  The most
luminous in $K_s$ is the object GL\,490.  Other objects brighter than
$K_s$\,=\,9 mag are SL\,82, SL\,131 and SL\,132 from Paper II and
SL\,171 from Table 1 (numbered in Figure 9).  The SED curves of these
objects are of Class II, and there is a high probability that their
masses correspond to A or B stars, i.e., they may belong to Herbig Ae/Be
stars.  Another alternative is that these stars are located in space
closer to the Sun than we accepted.

Other difference between the two SFRs is in the number of objects with
$H$--$K_s$\,$>$\,1.5 -- in Cam OB1 only GL\,490 falls in this color
range, while in Taurus such red objects are numerous, but at a lower
luminosity level.  No doubt, this effect originates in the accepted
accuracy limit of $H$ and $K_s$ magnitudes ($<$\,0.05 mag) in our study.
This effect cuts off all the red objects fainter than $K_s$ = 14 mag at
$H$--$K_s$ = 0.6 and fainter than $K_s$ = 12 mag at $H$--$K_s$ = 1.5.

\sectionb{8}{J--H, H--K$_{\rm s}$ DIAGRAM FOR THE CAM OB1 SFR}

In Paper II the $J$--$H$ vs.  $H$--$K_s$ diagram was given for real and
suspected young objects belonging to the Local arm.  Then only the YSOs
with $H$--$K_s$\,$\geq$\,1.0 were known.  Now we have 20 additional
suspected YSOs with $H$--$K_s$ between 0.5 and 1.0.  Thus, now we are
able to plot a more complete two-color diagram for the objects in the
Local arm.  Since in the present paper we concentrate on the Cam OB1
star-forming region, we will ignore the YSOs from Paper II found to
belong to the Gould Belt layer.


\begin{figure}[!th]
\centerline{\psfig{figure=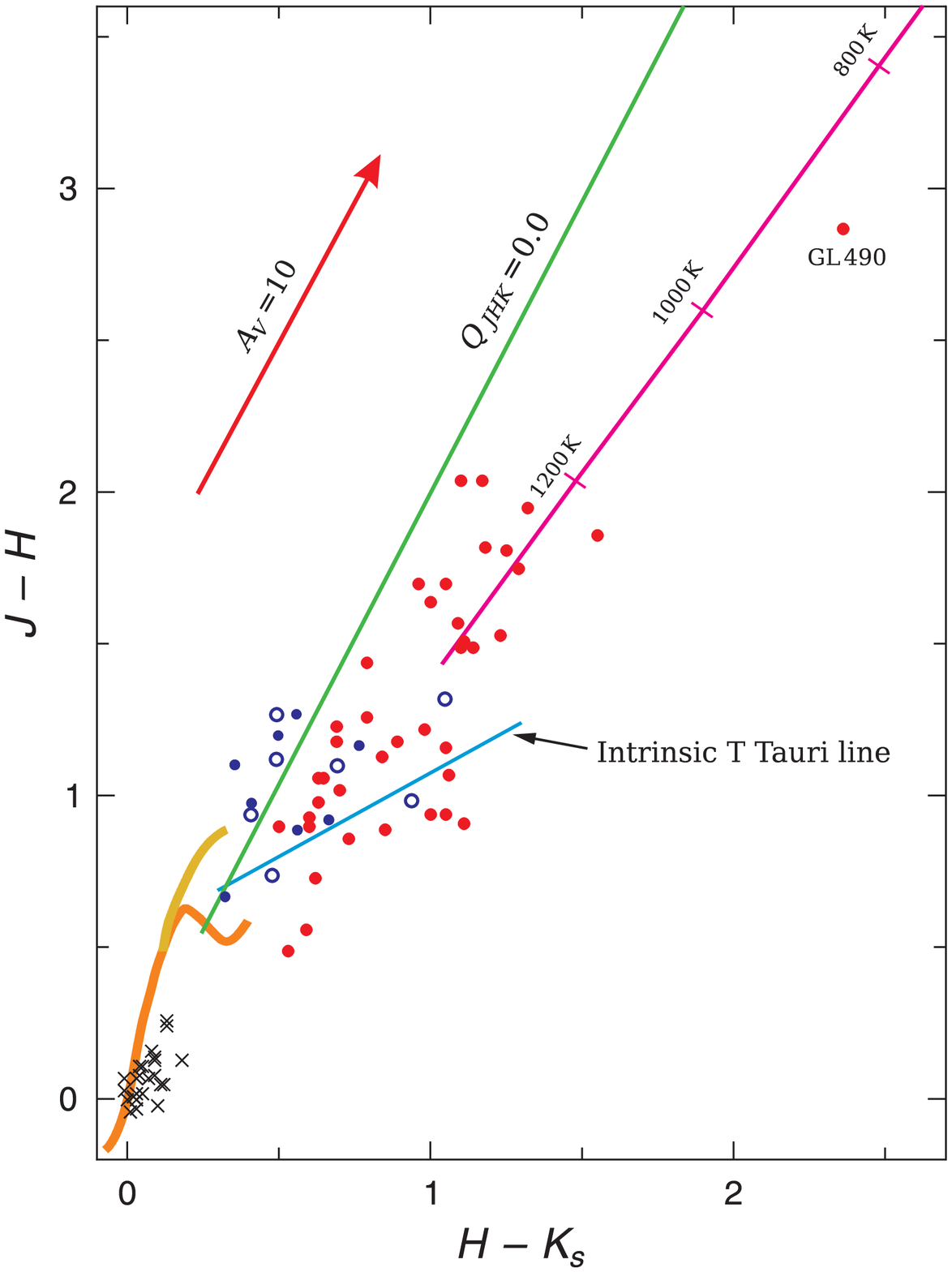,width=95mm,angle=0,clip=true}}
\vspace{.8mm}
\captionb{11}{The $J$--$H$ vs.  $H$--$K_s$ diagram for the suspected
YSOs (red dots) and other young objects in the Cam OB1 star-forming
region.  Black crosses are O--B3 stars of the Cam OB1 association, blue
dots designate known irregular variables and blue open circles designate
H$\alpha$ emission stars.  The intrinsic main-sequence and giant lines
are shown in orange and yellow.  The blue line designates the
intrinsic locus of T Tauri stars (Meyer et al. 1997), the violet line is
the locus of black bodies.  The length of the reddening vector (in red)
corresponds to the extinction in the $V$ passband of 10 mag.  Young
stellar objects were searched for in the region below the green line.}
\end{figure}

Figure 11 shows the $J$--$H$ vs.  $H$--$K_s$ diagram for the same YSOs
which were plotted in the color-magnitude diagram (Figure 9), but now
all of them are shown as red dots.  The dots between $H$--$K_s$ = 0.5
and 1.0 designate objects of the 3\,$\times$\,3\degr\ area around
GL\,490, while the dots with $H$--$K_s$\,$\geq$\,1.0 designate objects
in the whole 26\,$\times$\,24\degr\ area.  Additionally, we have plotted
OB stars of the association Cam OB1 (black crosses), known irregular
variables (blue dots) and the H$\alpha$ emission stars (blue circles).
Probably not all variables and emission-line stars belong to the Cam OB1
star-forming region, some may be located closer to the Sun.  Figure 11
also shows the loci of normal luminosity V stars (orange curve) and red
giants (yellow curve), the reddening vector for $A_V$ = 10 mag (red),
the reddening line of OB stars ($Q_{JHK}$ = 0.0, green), the black-body
line (violet) and the intrinsic line of T Tauri stars (blue).

It is evident, that the $J$--$H$ vs.  $H$--$K_s$ diagram for YSOs in the
Cam OB1 SFR is very similar to those in other SFRs.  However, some red
dots lie lower than the intrinsic line of T Tauri stars.  Probably some
of them (especially SL\,175, SL\,176 and SL\,183) are heavily reddened
Herbig Ae/Be stars or related objects in an earlier evolutionary stage.
These stars, like their counterparts above the intrinsic T Tauri line,
exhibit infrared excesses at $>$\,2 $\mu$m.

\sectionb{9}{CONCLUSIONS}

In the dust cloud DoH\,942, at the center of the Cam OB1 association, we
have identified 50 infrared objects suspected to be in the
pre-main-sequence stage of evolution.  The criteria for the attribution
of objects to YSOs were their positions in the $J$--$H$ vs.  $H$--$K_s$
diagram and their spectral energy distribution curves constructed using
the 2MASS, IRAS and MSX data.  Among the 25 objects with the IRAS and/or
MSX data we identify 16 YSOs of Class I, 8 YSOs of Class II and one
object may be heavily reddened Ae/Be star or a background AGB star.

The MSX data confirm that the objects, forming the upper `tail' in the
$J$--$H$ vs.  $H$--$K_s$ diagram, are background K--M giants heavily
reddened by the DoH\,942 dust cloud.  The comparison field in the nearby
area with a relatively low interstellar reddening does not contain
objects which might be suspected as pre-main-sequence stars.

The suspected YSOs in the color-magnitude plane $K_s$ vs.  $H$--$K_s$
occupy a large area located right of the main sequence, like YSOs in
other star-forming regions (Orion, Taurus, etc.).  This position can be
explained by their luminosity, interstellar and circumstellar reddening
and the infrared thermal emission from circumstellar envelopes and
disks.  However, the available data are not sufficient to disentangle
all these effects for individual stars.  However, some conclusions can
be drawn for the most luminous objects like GL\,490.  The other three
objects brighter than $K_s$ = 9 mag (and some fainter objects) probably
are heavily reddened Herbig Ae/Be stars or their prestellar
counterparts.

\thanks {We are thankful to Bo Reipurth for useful comments and to
Edmundas Mei\v stas and Stanislava Barta\v si\= ut\.e for their help
preparing the paper.  The use of the 2MASS, IRAS, MSX, SkyView and
Simbad databases is acknowledged.}

\References

\refb Bessell M. S., Brett J. M. 1988, PASP, 100, 1134

\refb Brice\~{n}o C., Luhman K. L., Hartmann L., Stauffer J. R.,
 Kirkpatrick J. D. 2002, ApJ, 580, 317

\refb Cardelli J. A., Clayton G. C., Mathis J. S. 1989, ApJ, 345, 245

\refb Cutri R. M., Skrutskie M. F., Van Dyk S., Beichman C. A. et al.
2003, {\it 2MASS All Sky Catalog of Point Sources}, NASA/IPAC Infrared
Science Archive,\\ http://irsa.ipac.caltech.edu/applications/Gator/

\refb Dobashi K., Uehara H., Kandori R., Sakurai T., Kaiden M.,
Umemoto T., Sato F. 2005, PASJ, 57, S1

\refb Hillenbrand L. A., Carpenter J. M. 2000, ApJ, 540, 236

\refb Hodapp K.-W. 1990, ApJ, 352, 184

\refb Hodapp K.-W. 1994, ApJS, 94, 615

\refb Lada C. J. 1987, in {\it Star Forming Regions} (IAU Symp. 115),
eds. M. Peimbert \& J. Jugaku, Reidel Publ. Comp., Dordrecht, p.\,1

\refb Meyer M. R., Calvert N., Hillenbrand L. A. 1997, AJ, 114, 288

\newpage

\refb Myers P. C., Fuller G. A., Mathieu R. D., Beichman C. A., Benson
P. J., Schild R. E., Emerson J. P. 1987, ApJ, 319, 340

\refb Robitaille T. P., Whitney B. A., Indebetouw R., Wood K., Denzmore
P. 2006, ApJS, 167, 256

\refb Robitaille T. P., Whitney B. A., Indebetouw R., Wood K. 2007,
ApJS, 169, 328

\refb Skrutskie M. F., Cutri R. M., Stiening R., Weinberg M. D. et al.
2006, AJ, 131, 1163

\refb Strai\v zys V., Laugalys V. 2007a, Baltic Astronomy, 16, 167
(Paper I)

\refb Strai\v zys V., Laugalys V. 2007b, Baltic Astronomy, 16, 327
(Paper II)

\refb Whitney B. A., Wood K., Bjorkman J. E., Wolff M. J.. 2003, ApJ,
591, 1049

\end{document}

%% file: table1.tex
\def\hdstrut{\vrule height13pt depth0pt width0pt}
\landscape
\noindent
\tabcolsep=9pt
\begin{longtable}{rcccccrcccl}
\multicolumn{11}{l}{\hspace{10mm}\parbox{120mm}{\baselineskip=9pt
{\normbf\ \ Table 1.}\vtop{{\norm\ Suspected YSOs with $H$--$K_s$
between 0.5 and 1.0 in the 3\degr\,$\times$\,3\degr\ area centered at
$\ell$, $b$\,=\,142.5\degr, +1.0\degr.  \lstrut}}}}\\
\hline
\noalign{\vskip1.5mm}
  SL  &   $\ell$  &  $b$   &  $F$ & $J$ & $H$ & $K_s$~~ &  $J$--$H$ &
$H$--$K_s$ & $Q_{JHK}$ & IRAS, MSX \\
\noalign{\vskip1.5mm}
\hline
\endfirsthead
\multicolumn{11}{l}{\hspace{10mm}{\normbf\ \ Table 1.}{\norm\ Continued\lstrut}}\\
\hline
\noalign{\vskip1.5mm}
  SL  &   $\ell$  &  $b$   &  $F$ & $J$ & $H$ & $K_s$~~ &
$J$--$H$ & $H$--$K_s$ & $Q_{JHK}$ & IRAS, MSX \\
\noalign{\vskip1.5mm}
\hline
\endhead
\endfoot
\noalign{\vskip1.5mm}
\noalign{Objects with $H$--$K_s$ between 0.75 and 1.00}
\noalign{\vskip.5mm}
 143 &   141.859 &  1.753 &  --    &  15.58 & 14.21 & 13.42 & 1.37 & 0.79 & --0.09 &  \\[-1.5pt]
 144 &   141.900 &  1.666 &  16.30 &  12.55 & 11.34 & 10.46 & 1.20 & 0.88 & --0.42 &  \\[-1.5pt]
 145 &   141.938 &  1.764 &  --    &  15.24 & 13.73 & 12.84 & 1.51 & 0.90 & --0.14 &  \\[-1.5pt]
 146 &   141.955 &  1.740 &  18.79 &  13.88 & 12.49 & 11.69 & 1.39 & 0.80 & --0.10 &  \\[-1.5pt]
 147 &   141.967 &  1.715 &  --    &  14.68 & 13.08 & 12.20 & 1.60 & 0.88 & --0.03 &  \\[-1.5pt]
 148 &   142.006 &  1.749 &  19.16 &  14.56 & 13.19 & 12.38 & 1.37 & 0.81 & --0.13 &  \\[-1.5pt]
 149\rlap{*} &   142.041 &  1.849 &  --    &  15.41 & 13.91 & 13.09 &  1.50 & 0.82 & --0.01 & \\[-1.5pt]
 150 &   142.164 &  1.670 &  --    &  15.30 & 13.77 & 12.94 & 1.53 & 0.84 & --0.02 &  \\[-1.5pt]
 151 &   142.243 &  1.740 &  --    &  15.16 & 13.74 & 12.95 & 1.42 & 0.79 & --0.04 &  \\[-1.5pt]
 152 &   142.265 &  1.693 &  --    &  15.02 & 13.63 & 12.80 & 1.39 & 0.82 & --0.13 &  \\[-1.5pt]
 153 &   142.299 &  0.619 &  16.51 &  11.98 & 10.99 & 10.23 & 1.00 & 0.76 & --0.42 &  \\[-1.5pt]
 154 &   142.341 &  1.592 &  19.38 &  14.34 & 12.96 & 12.20 & 1.37 & 0.77 & --0.05 &  \\[-1.5pt]
 155 &   142.403 &  1.220 &  18.56 &  14.76 & 13.58 & 12.83 & 1.18 & 0.75 & --0.20 &  \\[-1.5pt]
 156 &   142.477 &  1.683 &  --    &  15.48 & 13.98 & 13.02 & 1.50 & 0.96 & --0.27 &  \\[-1.5pt]
 157 &   142.529 &  1.577 &  --    &  15.52 & 13.97 & 13.12 & 1.56 & 0.85 & --0.01 &  \\[-1.5pt]
 158 &   142.580 &  1.539 &  14.63 &  11.59 & 10.37 &  9.38 & 1.22 & 0.98 & --0.60 &  03261+5803, MSX \\[-1.5pt]
 159 &   142.633 &  1.418 &  18.10 &  14.06 & 12.81 & 12.02 & 1.26 & 0.79 & --0.20 &  03260+5755 \\[-1.5pt]
 160 &   142.647 &  1.185 &  20.35 &  15.25 & 14.05 & 13.25 & 1.20 & 0.80 & --0.29 &  \\[-1.5pt]
 161 &   142.680 &  1.425 &  18.13 &  14.17 & 12.92 & 12.12 & 1.25 & 0.81 & --0.24 &  03260+5755 \\[-1.5pt]
 162\rlap{*} &   142.700 &  1.689 &  17.78 &  13.17 & 11.86 & 10.92 & 1.31 & 0.94  & --0.43 &  \\[-1.5pt]
 163 &   142.743 &  1.946 &  17.51 &  13.89 & 12.81 & 11.82 & 1.08 & 0.99 & --0.75 &  03289+5818 \\[-1.5pt]
 164\rlap{*} &   142.746 &  1.390 &  20.27 &  15.56 & 14.04 & 13.19 & 1.53 & 0.84 & --0.03 &  \\[-1.5pt]
 165\rlap{*} &   142.972 &  1.878 &  15.40 &  12.16 & 10.99 & 10.10 & 1.18 & 0.89 & --0.47 &  MSX \\[-1.5pt]
 166 &   142.981 &  1.707 &  19.79 &  15.45 & 14.26 & 13.43 & 1.19 & 0.83  & --0.35 &  \\[-1.5pt]
 167\rlap{*} &   142.990 &  1.751 &  18.61 &  14.32 & 12.95 & 12.18 & 1.38 & 0.77 & --0.05 &  \\[-1.5pt]
 168\rlap{*} &   142.993 &  1.755 &  18.66 &  13.82 & 12.45 & 11.59 & 1.38 & 0.86 & --0.21 &  \\[-1.5pt]
 169 &   143.004 &  1.753 &  18.72 &  14.60 & 13.16 & 12.37 & 1.44 & 0.79 & --0.02 &  03298+5758 \\[-1.5pt]
 170 &   143.008 &  1.429 &  18.84 &  15.15 & 13.93 & 13.17 & 1.22 & 0.76 & --0.19 &  \\[-1.5pt]
 171 &   143.224 &  0.941 &  16.60 &  10.03 &  8.33 &  7.37 & 1.70 & 0.96 & --0.08 & MSX\hdstrut  \\[-1.5pt]
 172 &   143.369 &  1.262 &  18.62 &  14.75 & 13.64 & 12.86 & 1.11 & 0.78 & --0.32 &  \\[-1.5pt]
 173 &   143.908 &  0.644 &  13.63 &  11.78 & 10.89 & 10.03 & 0.89 & 0.85 & --0.69 &  03304+5633 \\[-1.5pt]
 174\rlap{*} &   143.958 &--0.458 &  19.65 &  15.34 & 14.21 & 13.37 & 1.13 & 0.84 & --0.42  &  03262+5536 \\[-1.5pt]
\noalign{\vskip.5mm}
\noalign{Objects with $H$--$K_s$ between 0.50 and 0.75}
\noalign{\vskip.5mm}
 175\rlap{*} &     141.245 &  1.375 &  13.43 &  11.72 & 11.23 & 10.70 & 0.49 & 0.53 & --0.50 & 03167+5840 \\[-1.5pt]
 176\rlap{*} &     141.353 &  0.351 &  14.49 &  12.40 & 11.66 & 11.04 & 0.73 & 0.62 & --0.42 & 03135+5743 \\[-1.5pt]
 177 &     141.927 &  2.004 &  16.38 &  13.16 & 12.13 & 11.44 & 1.02 & 0.70 & --0.27 & 03239+5849 \\[-1.5pt]
 178 &     141.988 &  0.964 &  19.91 &  14.70 & 13.65 & 13.02 & 1.06 & 0.63 & --0.10 & 03199+5755 \\[-1.5pt]
 179 &     142.005 &  0.839 &   --   &  15.38 & 14.46 & 13.85 & 0.93 & 0.60 & --0.18 & 03194+5746 \\[-1.5pt]
 180 &     142.075 &  1.163 &  15.06 &  12.05 & 11.19 & 10.45 & 0.86 & 0.73 & --0.49 & 03213+5801 \\[-1.5pt]
 181\rlap{*} &     142.457 &  1.722 &  19.77 &  15.22 & 14.32 & 13.71 & 0.90 & 0.60 & --0.21 & 03263+5816 \\[-1.5pt]
 182 &     142.553 &  0.847 &  20.35 &  15.30 & 14.31 & 13.68 & 0.98 & 0.63 & --0.18 & 03231+5730 \\[-1.5pt]
 183\rlap{*} &     142.604 &  1.201 &  14.02 &  12.36 & 11.81 & 11.22 & 0.56 & 0.59 & --0.53 & 03248+5745 \\[-1.5pt]
 184 &     142.650 &  1.428 &  17.28 &  13.02 & 11.80 & 11.11 & 1.23 & 0.69 & --0.05 & 03260+5755 \\[-1.5pt]
 185\rlap{*} &     142.788 &  1.528 &  20.50 &  15.05 & 13.99 & 13.35 & 1.06 & 0.64 & --0.12 & 03275+5755 \\[-1.5pt]
 186 &     142.868 &  1.818 &  18.61 &  14.55 & 13.38 & 12.69 & 1.18 & 0.69 & --0.09 & 03292+5806 \\[-1.5pt]
 187 &     143.644 &  1.322 &  18.33 &  14.15 & 13.25 & 12.75 & 0.90 & 0.50 & --0.02 & 03317+5716 \\[+.2pt]
\hline
\end{longtable}
\vspace{-2mm}
\enlargethispage{3mm}

{\bf Notes to Table 1}:
\vskip1mm

 SL\,149:  binary, in $K$ brighter component at 6\,\arcsec;
\vspace{-.3mm}

 SL\,162:  binary, in $K$ fainter component at 6\,\arcsec;
\vspace{-.3mm}

 SL\,164:  no object in $R$ and very faint in $K$. Wrong coordinates?
\vspace{-.3mm}

 SL\,165:  in $R$ with tail, in $K$ no tail;
\vspace{-.3mm}

 SL\,167 and 168:  nearby YSOs separated by 19\,\arcsec;
\vspace{-.3mm}

 SL\,174:  very faint in $R$ and $K$. Wrong coordinates and IRAS
identification?
\vspace{-.3mm}

 SL\,175, 176 and 183:  probably heavily reddened Herbig Ae/Be stars
 (in the $J$--$H$ vs. $H$--$K_s$ diagram they lie well\\
 \hhuad\hhuad\hhuad\hhuad\hhuad below the intrinsic T Tauri line);
\vspace{-.3mm}

 SL\,181:  binary, in $K$ fainter component at 10\,\arcsec;
\vspace{-.3mm}

 SL\,185:  binary, in $K$ fainter component at 12\,\arcsec.
\endlandscape

%% file: table2.tex
\begin{table}[!t]
\begin{center}
\vbox{\small\tabcolsep=6pt
\parbox[c]{120mm}{\baselineskip=10pt
{\normbf\ \ Table 2.}{\small\ IRAS and MSX data for the suspected YSOs in the 
investigated area with $H$--$K_s$\,$\geq$\,0.75. Fluxes are given in Janskys.\lstrut}}
\begin{tabular}{rcrrrrrc}
\tablerule
 ~SL &    IRAS   &  $F_{12}$ & $F_{25}$ & $F_{60}$ & $F_{100}$ & $F_{8.3}$ & YSO type \\
\tablerule
 89 & 03228+5834 &   0.24: &   1.06 &   $<$4.04  &   $<$27.7  &    --~~   &    I     \\
 93 & 03233+5833 &   0.44  &   2.13 &   23.3   &    40\rlap{:}   &    --~~   &    I         \\
 95 & 03236+5836 &  90.5   &   290  &    715   &   1156   &  54.73  &    I         \\
102 & 03290+5724 &  $<$0.26  &   0.17 &   $<$0.75  &   $<$44.76 &    --~~   &    II  \\
107 & 03303+5643 &  $<$0.39  &  $<$0.25 &    0.69  &   $<$5.80  &    --~~   &    I   \\
158 & 03261+5803 &   0.33  &   0.32 &   $<$3.87  &   $<$44.34 &   0.25  &    II    \\
159 & 03260+5755 &  $<$0.31  &   0.25 &   $<$3.89  &   $<$57.19 &    --~~   &    II  \\
165 & --     &  --~~    & --~~   & --~~    &  --~~   &   0.14  &    II        \\
169 & 03298+5758 &   0.40  &   0.59 &    6.32  &    26.86 &    --~~   &    I         \\
171 & --     &  --~~    & --~~   & --~~    &  --~~   &   0.39  &    II        \\
173 & 03304+5633 &  $<$0.49  &   0.31 &   $<$0.63  &    $<$5.38 &    --~~   &        \\
174 & 03262+5536 &  $<$0.25  &  $<$0.25 &    1.15  &     8.35 &    --~~   &    I     \\
175 & 03167+5840 &   0.75  &   0.34 &    1.51\rlap{:} &   $<$26.32 & --~~   &    I:      \\
176 & 03135+5743 &  $<$0.57  &   0.20 &   $<$3.26  &   $<$25.98 &    --~~   &    II  \\
177 & 03239+5849 &  $<$0.44  &  $<$0.36 &    5.93  &   $<$33.39 &    --~~   &    I   \\
178 & 03199+5755 &  $<$0.37  &  $<$0.25 &   $<$2.01  &    16.05 &    --~~   &    I   \\
179 & 03194+5746 &  $<$0.28  &  $<$0.27 &    0.83  &   $<$27.62 &    --~~   &    I   \\
180 & 03213+5801 &  $<$0.29  &   0.34 &   $<$2.54  &   $<$33.80 &    --~~   &    II  \\
181 & 03263+5816 &  $<$0.44  &  $<$0.26 &    1.29  &   $<$48.02 &    --~~   &    I   \\
182 & 03231+5730 &  $<$0.25  &  $<$0.25 &    0.57\rlap{:} &     9.12 &    --~~   &    I     \\
183 & 03248+5745 &   0.31\rlap{:} &   0.96 &   $<$2.94  &   $<$28.79 & --~~   &    I     \\
184 & 03260+5755 &  $<$0.31  &   0.25 &   $<$3.89  &   $<$57.19 &    --~~   &    II  \\
185 & 03275+5755 &  $<$0.29  &  $<$0.31 &   $<$2.32  &     8.46 &    --~~   &    I   \\
186 & 03292+5806 &  $<$0.28  &   0.43 &   $<$3.35  &    10.60 &    --~~   &    I     \\
187 & 03317+5716 &  $<$0.30  &  $<$0.25 &    0.59  &   $<$33.54 &    --~~   &    I   \\
\tablerule
\end{tabular}
}
\end{center}
\vspace{-5mm}
\end{table}

%% file: table3.tex
\begin{center}
\small
\noindent
\tabcolsep=6pt
\begin{longtable}{lcrrrrrcc}
\multicolumn{9}{l}{\parbox{105mm}{\baselineskip=9pt
{\normbf\ \ Table 3.}{\norm\ Suspected heavily reddened K--M III stars
with\\ $H$--$K_s$\,$\geq$\,0.75 in the 3\degr\,$\times$\,3\degr\ area
centered at $\ell$, $b$\,=\,142.5\degr, +1.0\degr.\lstrut}}}\\
\hline
\noalign{\vskip1.5mm}
K-M  & $\ell$ &   $b$~~~   &  $J$~~~  & $H$~~   & $K_s$~ & $J$--$H$ &
$H$--$K_s$ & $Q_{JHK}$  \\
\noalign{\vskip1mm}
\hline
\noalign{\vskip1mm}
\endfirsthead
\multicolumn{9}{l}{{\normbf\ \ Table 3.}{\norm\ Continued\lstrut}}\\
\hline
\noalign{\vskip1.5mm}
K-M  & $\ell$ &    $b$~~~ &  $J$~~~  & $H$~~ & $K_s$~ & $J$--$H$ &
$H$--$K_s$ & $Q_{JHK}$  \\
\noalign{\vskip1mm}
\hline
\noalign{\vskip1mm}
\endhead
\endfoot
01 &  141.775 &   1.886 &   14.77 &  12.93 &  12.11 &  1.84 &  0.82 &  0.32   \\
02 &  141.838 &   1.232 &   13.00 &  10.28 &   9.03 &  2.72 &  1.25 &  0.40   \\
03 &  141.896 &   0.473 &   15.12 &  13.22 &  12.37 &  1.89 &  0.86 &  0.31   \\
04 &  141.908 &   0.482 &   14.32 &  12.63 &  11.88 &  1.68 &  0.75 &  0.29   \\
05 &  141.916 &   1.869 &   12.33 &  10.45 &   9.64 &  1.88 &  0.81 &  0.39   \\
06 &  141.918 &   1.855 &   13.80 &  12.12 &  11.35 &  1.68 &  0.76 &  0.27   \\
07 &  141.925 &   1.756 &   13.98 &  11.85 &  10.84 &  2.14 &  1.01 &  0.26   \\
08 &  141.930 &   1.747 &   13.00 &  10.76 &   9.71 &  2.23 &  1.05 &  0.28   \\
09 &  141.935 &   1.115 &   12.23 &  10.27 &   9.44 &  1.96 &  0.83 &  0.42   \\
10 &  141.939 &   1.158 &   14.34 &  12.46 &  11.66 &  1.88 &  0.80 &  0.41   \\
11 &  141.943 &   0.498 &   13.96 &  12.07 &  11.24 &  1.89 &  0.83 &  0.36   \\
12 &  141.944 &   0.538 &   13.52 &  11.84 &  11.08 &  1.68 &  0.76 &  0.27   \\
13 &  141.945 &   1.702 &   14.22 &  12.46 &  11.69 &  1.76 &  0.78 &  0.32   \\
14 &  141.951 &   1.185 &   14.49 &  12.59 &  11.80 &  1.91 &  0.78 &  0.46   \\
15 &  141.975 &   1.222 &   10.56 &   8.78 &   8.03 &  1.78 &  0.76 &  0.38   \\
16 &  141.983 &   1.856 &   11.00 &   9.09 &   8.20 &  1.90 &  0.89 &  0.26   \\
17 &  141.993 &   1.694 &   14.65 &  12.80 &  11.98 &  1.84 &  0.82 &  0.33   \\
18 &  141.997 &   1.008 &   15.31 &  13.44 &  12.64 &  1.86 &  0.80 &  0.38   \\
19 &  142.003 &   0.609 &   15.46 &  13.83 &  13.07 &  1.63 &  0.76 &  0.22   \\
20 &  142.013 &   1.722 &   15.41 &  13.28 &  12.29 &  2.12 &  1.00 &  0.28   \\
21 &  142.031 &  1.823  &   13.91 &  11.03 &   9.66 &  2.88 &  1.37 &  0.34   \\
22 &  142.035 &  1.916  &   15.11 &  13.34 &  12.58 &  1.74 &  0.79 &  0.29   \\
23 &  142.041 &  1.807  &   13.87 &  12.02 &  11.18 &  1.85 &  0.84 &  0.31   \\
24 &  142.041 &  0.556  &   14.91 &  12.88 &  11.98 &  2.02 &  0.91 &  0.34   \\
25 &  142.045 &  1.045  &   14.56 &  12.88 &  12.12 &  1.69 &  0.75 &  0.29   \\
26 &  142.061 &  1.807  &   14.10 &  12.22 &  11.38 &  1.88 &  0.83 &  0.34   \\
27 &  142.067 &  1.068  &   12.04 &  10.11 &   9.31 &  1.93 &  0.79 &  0.46   \\
28 &  142.069 &  2.185  &   14.64 &  12.47 &  11.47 &  2.17 &  1.01 &  0.30   \\
29 &  142.071 &  1.635  &   14.59 &  12.72 &  11.87 &  1.87 &  0.85 &  0.29   \\
30 &  142.072 &  1.766  &   13.99 &  11.98 &  11.04 &  2.01 &  0.94 &  0.28   \\
31 &  142.083 &  1.619  &   12.19 &  10.21 &   9.36 &  1.98 &  0.85 &  0.42   \\
32 &  142.083 &  1.720  &   15.09 &  13.14 &  12.24 &  1.95 &  0.90 &  0.28   \\
33 &  142.089 &  0.547  &   15.05 &  13.22 &  12.37 &  1.83 &  0.85 &  0.25   \\
34 &  142.089 &  1.733  &   14.73 &  12.71 &  11.74 &  2.02 &  0.97 &  0.22   \\
35 &  142.101 &  1.682  &   13.35 &  11.62 &  10.86 &  1.73 &  0.76 &  0.31   \\
36 &  142.103 &  1.636  &   14.33 &  12.54 &  11.70 &  1.79 &  0.84 &  0.23   \\
37 &  142.105 &  0.711  &   12.85 &  11.02 &  10.27 &  1.82 &  0.75 &  0.43   \\
38 &  142.106 &  0.603  &   15.44 &  13.41 &  12.52 &  2.02 &  0.89 &  0.38   \\
39 &  142.109 &  1.653  &   14.02 &  11.88 &  10.93 &  2.14 &  0.96 &  0.37   \\
40 &  142.109 &  1.545  &   15.06 &  13.38 &  12.62 &  1.68 &  0.76 &  0.27   \\
41 &  142.116 &  1.727  &   15.10 &  13.42 &  12.66 &  1.68 &  0.76 &  0.28   \\
42 &  142.123 &  1.647  &   11.67 &   9.40 &   8.37 &  2.27 &  1.03 &  0.36   \\
43 &  142.126 &  1.834  &   11.79 &   9.96 &   9.21 &  1.83 &  0.76 &  0.43   \\
44 &  142.134 &  1.688  &   15.31 &  13.33 &  12.48 &  1.98 &  0.85 &  0.40   \\
45 &  142.141 &  0.599  &   14.30 &  12.51 &  11.70 &  1.79 &  0.81 &  0.28   \\
46 &  142.145 &  0.558  &   15.56 &  13.19 &  12.07 &  2.37 &  1.12 &  0.30   \\
47 &  142.158 &  0.759  &   13.55 &  11.39 &  10.49 &  2.16 &  0.90 &  0.49   \\
48 &  142.161 &  1.729  &   13.89 &  11.96 &  11.14 &  1.93 &  0.82 &  0.41   \\
49 &  142.173 &  1.579  &   15.18 &  13.19 &  12.32 &  1.99 &  0.87 &  0.38   \\
50 &  142.175 &  0.600  &   11.55 &   9.79 &   9.04 &  1.76 &  0.75 &  0.37   \\
51 &  142.180 &  0.628  &   11.07 &   8.62 &   7.48 &  2.45 &  1.14 &  0.35   \\
52 &  142.182 &  1.742  &   15.23 &  13.50 &  12.68 &  1.73 &  0.82 &  0.22   \\
53 &  142.191 &  1.651  &   14.75 &  12.41 &  11.29 &  2.34 &  1.12 &  0.26   \\
54 &  142.200 &  1.730  &   13.30 &  11.33 &  10.46 &  1.96 &  0.87 &  0.36   \\
55 &  142.207 &  1.750  &   14.86 &  13.24 &  12.49 &  1.62 &  0.75 &  0.23   \\
56 &  142.207 &  0.624  &   14.47 &  12.72 &  11.94 &  1.75 &  0.78 &  0.31   \\
57 &  142.219 &  1.671  &   13.34 &  11.42 &  10.56 &  1.93 &  0.85 &  0.35   \\
58 &  142.223 &  1.708  &   12.89 &  11.05 &  10.29 &  1.83 &  0.76 &  0.42   \\
59 &  142.255 &  1.644  &    8.68 &   6.76 &   5.87 &  1.91 &  0.90 &  0.26   \\
60 &  142.295 &  0.003  &    9.44 &   7.61 &   6.74 &  1.84 &  0.87 &  0.23   \\
61 &  142.345 &  1.735  &   11.47 &   9.67 &   8.90 &  1.80 &  0.77 &  0.37   \\
62 &  142.392 &  1.599  &   11.23 &   9.34 &   8.48 &  1.89 &  0.86 &  0.30   \\
63 &  142.401 &  1.158  &   14.74 &  12.75 &  11.88 &  1.99 &  0.87 &  0.38   \\
64 &  142.494 &  1.386  &   10.89 &   9.18 &   8.42 &  1.71 &  0.76 &  0.31   \\
65 &  142.501 &  1.554  &   12.96 &  10.20 &   8.89 &  2.76 &  1.31 &  0.34   \\
66 &  142.519 &  1.341  &   14.51 &  12.26 &  11.28 &  2.26 &  0.98 &  0.45   \\
67 &  142.552 &  1.406  &   14.27 &  12.20 &  11.31 &  2.07 &  0.89 &  0.42   \\
68 &  142.571 &  0.589  &   12.08 &  10.30 &   9.52 &  1.78 &  0.77 &  0.36   \\
69 &  142.572 &  1.424  &   15.17 &  13.50 &  12.73 &  1.67 &  0.76 &  0.26   \\
70 &  142.601 &  1.758  &   13.24 &  11.36 &  10.58 &  1.88 &  0.77 &  0.46   \\
71 &  142.630 &  1.610  &   13.96 &  11.84 &  10.93 &  2.12 &  0.91 &  0.43   \\
72 &  142.674 &  1.680  &   13.49 &  11.69 &  10.89 &  1.81 &  0.79 &  0.34   \\
73 &  142.682 &  0.480  &   12.75 &  10.81 &   9.98 &  1.94 &  0.83 &  0.41   \\
74 &  142.689 &  1.772  &   13.71 &  11.80 &  10.92 &  1.92 &  0.88 &  0.29   \\
75 &  142.706 &  1.656  &   12.47 &  10.70 &   9.94 &  1.77 &  0.76 &  0.37   \\
76 &  142.716 &  1.602  &   11.51 &   9.65 &   8.82 &  1.87 &  0.83 &  0.33   \\
77 &  142.720 &  1.416  &   14.57 &  12.67 &  11.86 &  1.90 &  0.81 &  0.41   \\
78 &  142.726 &  1.548  &   12.18 &  10.36 &   9.60 &  1.83 &  0.76 &  0.43   \\
79 &  142.730 &  1.763  &   14.20 &  12.33 &  11.51 &  1.86 &  0.82 &  0.35   \\
80 &  142.800 &  1.412  &   14.74 &  12.97 &  12.21 &  1.77 &  0.76 &  0.37   \\
81 &  142.934 &  1.819  &    9.48 &   7.75 &   6.96 &  1.74 &  0.79 &  0.27   \\
82 &  143.054 &  1.728  &   13.48 &  11.22 &  10.17 &  2.26 &  1.05 &  0.32   \\
83 &  143.061 &  1.736  &   12.97 &  11.00 &  10.09 &  1.97 &  0.91 &  0.29   \\
84 &  143.188 &  1.719  &    9.84 &   8.09 &   7.30 &  1.75 &  0.79 &  0.29   \\
85 &  143.323 & --0.047  &   13.04 &  11.27 &  10.47 &  1.77 &  0.80 &  0.29   \\
86 &  143.411 &  1.520  &   13.22 &  11.27 &  10.47 &  1.96 &  0.80 &  0.47   \\
87 &  143.499 &  1.711  &   12.81 &  10.94 &  10.15 &  1.86 &  0.79 &  0.40   \\
88 &  143.604 &  1.267  &   10.46 &   8.70 &   7.93 &  1.76 &  0.76 &  0.35   \\
\hline
\end{longtable}
\end{center}

%% file: table4.tex
\begin{table}[!th]
\begin{center}
\vbox{\small\tabcolsep=6pt
\parbox[c]{120mm}{\baselineskip=10pt
{\normbf\ \ Table 4.}{\small\ Stars of the GL\,490 region located in the $J$--$H$ vs.
$H$--$K_s$ diagram in the upper `tail' and measured by MSX.\lstrut}}
\begin{tabular}{cccccc}
\tablerule
K-M & $\ell$ &  $b$ &  RA (J2000) &  DEC (J2000)  &   Flux at 8.3 $\mu$m \\   &  deg   &  deg & \kern-2mm h~~~m~~~s   &    $\circ~~~\prime~~~\prime\prime$ & Jy\\
\tablerule
51 &  142.180 &  0.628 & 03~23~45.33 &   +57~41~30.4  &      0.14\\
59 &  142.255 &  1.644 & 03~28~30.37 &   +58~29~37.6  &      0.58\\
60 &  142.295 &  0.003 & 03~21~56.11 &   +57~06~20.3  &      0.39\\
81 &  142.934 &  1.819 & 03~33~30.89 &   +58~15~00.7  &      0.15\\
84 &  143.188 &  1.719 & 03~34~38.41 &   +58~01~17.6  &      0.13\\
\tablerule
\end{tabular}
}
\end{center}
\vspace{-3mm}
\end{table}